\PassOptionsToPackage{unicode}{hyperref}
\PassOptionsToPackage{hyphens}{url}
\PassOptionsToPackage{dvipsnames,svgnames,x11names}{xcolor}
\documentclass[
  letterpaper,
  DIV=11,
  numbers=noendperiod]{scrartcl}

\usepackage{amsmath,amssymb}
\usepackage{lmodern}
\usepackage{setspace}
\usepackage{iftex}
\ifPDFTeX
  \usepackage[T1]{fontenc}
  \usepackage[utf8]{inputenc}
  \usepackage{textcomp} 
\else 
  \usepackage{unicode-math}
  \defaultfontfeatures{Scale=MatchLowercase}
  \defaultfontfeatures[\rmfamily]{Ligatures=TeX,Scale=1}
\fi
\IfFileExists{upquote.sty}{\usepackage{upquote}}{}
\IfFileExists{microtype.sty}{
  \usepackage[]{microtype}
  \UseMicrotypeSet[protrusion]{basicmath} 
}{}
\makeatletter
\@ifundefined{KOMAClassName}{
  \IfFileExists{parskip.sty}{%
    \usepackage{parskip}
  }{
    \setlength{\parindent}{0pt}
    \setlength{\parskip}{6pt plus 2pt minus 1pt}}
}{
  \KOMAoptions{parskip=half}}
\makeatother
\usepackage{xcolor}
\ifx\paragraph\undefined\else
  \let\oldparagraph\paragraph
  \renewcommand{\paragraph}[1]{\oldparagraph{#1}\mbox{}}
\fi
\ifx\subparagraph\undefined\else
  \let\oldsubparagraph\subparagraph
  \renewcommand{\subparagraph}[1]{\oldsubparagraph{#1}\mbox{}}
\fi

\usepackage{color}
\usepackage{fancyvrb}

\DefineVerbatimEnvironment{Highlighting}{Verbatim}{commandchars=\\\{\}}
\usepackage{framed}
\definecolor{shadecolor}{RGB}{241,243,245}
\newenvironment{Shaded}{\begin{snugshade}}{\end{snugshade}}

\newcommand{\AttributeTok}[1]{\textcolor[rgb]{0.40,0.45,0.13}{#1}}

\newcommand{\CommentTok}[1]{\textcolor[rgb]{0.37,0.37,0.37}{#1}}

\newcommand{\DecValTok}[1]{\textcolor[rgb]{0.68,0.00,0.00}{#1}}
\newcommand{\DocumentationTok}[1]{\textcolor[rgb]{0.37,0.37,0.37}{\textit{#1}}}

\newcommand{\FloatTok}[1]{\textcolor[rgb]{0.68,0.00,0.00}{#1}}
\newcommand{\FunctionTok}[1]{\textcolor[rgb]{0.28,0.35,0.67}{#1}}

\newcommand{\NormalTok}[1]{\textcolor[rgb]{0.00,0.23,0.31}{#1}}

\newcommand{\OtherTok}[1]{\textcolor[rgb]{0.00,0.23,0.31}{#1}}

\newcommand{\SpecialCharTok}[1]{\textcolor[rgb]{0.37,0.37,0.37}{#1}}

\newcommand{\StringTok}[1]{\textcolor[rgb]{0.13,0.47,0.30}{#1}}

\usepackage{longtable,booktabs,array}
\usepackage{calc} 
\usepackage{etoolbox}
\makeatletter
\patchcmd\longtable{\par}{\if@noskipsec\mbox{}\fi\par}{}{}
\makeatother
\IfFileExists{footnotehyper.sty}{\usepackage{footnotehyper}}{\usepackage{footnote}}
\makesavenoteenv{longtable}
\usepackage{graphicx}
\makeatletter
\def\maxwidth{\ifdim\Gin@nat@width>\linewidth\linewidth\else\Gin@nat@width\fi}
\def\maxheight{\ifdim\Gin@nat@height>\textheight\textheight\else\Gin@nat@height\fi}
\makeatother
\setkeys{Gin}{width=\maxwidth,height=\maxheight,keepaspectratio}
\makeatletter
\def\fps@figure{htbp}
\makeatother
\newlength{\cslhangindent}
\newlength{\csllabelwidth}
\newlength{\cslentryspacingunit} 
\newenvironment{CSLReferences}[2] 
 {
  \setlength{\parindent}{0pt}
  \ifodd #1
  \let\oldpar\par
  \def\par{\hangindent=\cslhangindent\oldpar}
  \fi
  \setlength{\parskip}{#2\cslentryspacingunit}
 }%
 {}
\usepackage{calc}

\usepackage{booktabs}
\usepackage{longtable}
\usepackage{array}
\usepackage{multirow}
\usepackage{wrapfig}
\usepackage{float}
\usepackage{colortbl}
\usepackage{pdflscape}
\usepackage{tabu}
\usepackage{threeparttable}
\usepackage{threeparttablex}
\usepackage[normalem]{ulem}
\usepackage{makecell}
\usepackage{xcolor}
\KOMAoption{captions}{tableheading}
\makeatletter
\@ifpackageloaded{caption}{}{\usepackage{caption}}
\AtBeginDocument{%
\ifdefined\contentsname
  \renewcommand*\contentsname{Table of contents}
\else
  \newcommand\contentsname{Table of contents}
\fi
\ifdefined\listfigurename
  \renewcommand*\listfigurename{List of Figures}
\else
  \newcommand\listfigurename{List of Figures}
\fi
\ifdefined\listtablename
  \renewcommand*\listtablename{List of Tables}
\else
  \newcommand\listtablename{List of Tables}
\fi
\ifdefined\figurename
  \renewcommand*\figurename{Figure}
\else
  \newcommand\figurename{Figure}
\fi
\ifdefined\tablename
  \renewcommand*\tablename{Table}
\else
  \newcommand\tablename{Table}
\fi
}
\@ifpackageloaded{float}{}{\usepackage{float}}
\floatstyle{ruled}
\@ifundefined{c@chapter}{\newfloat{codelisting}{h}{lop}}{\newfloat{codelisting}{h}{lop}[chapter]}
\floatname{codelisting}{Listing}

\makeatother
\makeatletter
\@ifpackageloaded{caption}{}{\usepackage{caption}}
\@ifpackageloaded{subcaption}{}{\usepackage{subcaption}}
\makeatother
\makeatletter
\@ifpackageloaded{tcolorbox}{}{\usepackage[many]{tcolorbox}}
\makeatother
\makeatletter
\@ifundefined{shadecolor}{\definecolor{shadecolor}{rgb}{.97, .97, .97}}
\makeatother
\ifLuaTeX
  \usepackage{selnolig}  
\fi
\IfFileExists{bookmark.sty}{\usepackage{bookmark}}{\usepackage{hyperref}}
\IfFileExists{xurl.sty}{\usepackage{xurl}}{} 
\hypersetup{
  pdftitle={Causal inference is not just a statistics problem},
  pdfauthor={Lucy D'Agostino McGowan; Travis Gerke; Malcolm Barrett},
  colorlinks=true,
  linkcolor={blue},
  filecolor={Maroon},
  citecolor={Blue},
  urlcolor={Blue},
  pdfcreator={LaTeX via pandoc}}

\title{Causal inference is not just a statistics problem}
\author{Lucy D'Agostino McGowan \and Travis Gerke \and Malcolm Barrett}
\date{}

\begin{document}
\maketitle
\ifdefined\Shaded\renewenvironment{Shaded}{\begin{tcolorbox}[interior hidden, enhanced, frame hidden, sharp corners, breakable, borderline west={3pt}{0pt}{shadecolor}, boxrule=0pt]}{\end{tcolorbox}}\fi

\setstretch{2}
\hypertarget{abstract}{%
\subsection{Abstract}\label{abstract}}

This paper introduces a collection of four data sets, similar to
Anscombe's Quartet, that aim to highlight the challenges involved when
estimating causal effects. Each of the four data sets is generated based
on a distinct causal mechanism: the first involves a collider, the
second involves a confounder, the third involves a mediator, and the
fourth involves the induction of M-Bias by an included factor. The paper
includes a mathematical summary of each data set, as well as directed
acyclic graphs that depict the relationships between the variables.
Despite the fact that the statistical summaries and visualizations for
each data set are identical, the true causal effect differs, and
estimating it correctly requires knowledge of the data-generating
mechanism. These example data sets can help practitioners gain a better
understanding of the assumptions underlying causal inference methods and
emphasize the importance of gathering more information beyond what can
be obtained from statistical tools alone. The paper also includes R code
for reproducing all figures and provides access to the data sets
themselves through an R package named quartets.

\hypertarget{introduction}{%
\subsection{Introduction}\label{introduction}}

This paper focuses on introducing causal inference concepts to students.
Many statistics courses incorporate related concepts such as the
difference between an observational study and an experiment, the use of
random assignment in experiments, and the power of paired data. The
focus here is specifically on how to select which variables to adjust
for when handling observational data, that is, data with non-randomized
exposure(s), when the goal is to estimate a causal effect. In a causal
inference setting, variable selection techniques meant for prediction
are often not appropriate; rather, we often rely on domain expertise and
a philosophical understanding of the interrelationship between measured
(and unmeasured) factors and the exposure and outcome of interest. The
following material is designed for students with basic training in
statistical modeling (i.e.~the ability to fit and interpret an ordinary
least squares regression model) and basic summary statistics, such as
correlation. In our experience using the following material in the
classroom, some students already knew about concepts covered (e.g.,
colliders, confounders, mediators, and M-bias), while others learned
about them for the first time. We often use a mix of theoretical
discussions and real-life examples to help students understand the
concepts better. The material discussed in this paper was created
specifically to give students the ability to closely examine data sets
that clearly demonstrate, as the paper title suggests, that \emph{causal
inference is not just a statistical problem}. These hands-on data sets
bring this statement out of the theoretical and into reality.

Anscombe's quartet is a set of four data sets with the same summary
statistics (means, variances, correlations, and linear regression fits)
but which exhibit different distributions and relationships when plotted
on a graph (Anscombe 1973). Often used to teach introductory statistics
courses, Anscombe created the quartet to illustrate the importance of
visualizing data before drawing conclusions based on statistical
analyses alone. Here, we propose a different quartet, where statistical
summaries do not provide insight into the underlying mechanism, but even
visualizations do not solve the issue. In these examples, an
understanding or assumption of the data-generating mechanism is required
to capture the relationship between the available factors correctly.
This proposed quartet can help practitioners better understand the
assumptions underlying causal inference methods, further driving home
the point that we require more information than can be gleaned from
statistical tools alone to estimate causal effects accurately.

The \emph{causal quartet} data sets presented in this paper are
available in an R package titled \texttt{quartets} (D'Agostino McGowan
2023). This package also includes other helpful data sets for teaching,
including Anscombe's quartet, the ``Datasaurus Dozen'' (Matejka and
Fitzmaurice 2017), an exploration of varying interaction effects (Rohrer
and Arslan 2021), a quartet of model types fit to the same data that
yield the same performance metrics but fit very different underlying
mechanisms (Biecek, Baniecki, and Krzyznski 2023), and a set of
conceptual causal quartets that highlight the impact of treatment
heterogeneity on the average treatment effect (Gelman, Hullman, and
Kennedy 2023).

\hypertarget{methods}{%
\subsection{Methods}\label{methods}}

We begin this section with a causal inference primer, including
reference to several commonly used terms as well as a description of the
assumptions needed when estimating causal effects using traditional
methodology, as suggested here. We then provide a primer in causal
diagrams, useful tools for communicating proposed causal relationships
between factors. This is followed by a description of the \emph{causal
quartet}, data sets intended to illustrate that causal inference is not
just a statistics problem. Finally, we describe the solution to this
proposed problem.

\hypertarget{sec-prim}{%
\subsubsection{Causal inference primer}\label{sec-prim}}

In causal inference, we are often trying to estimate the effect of some
exposure, \(X\), on some outcome \(Y\). One framework we use to think
through this problem is the ``potential outcomes'' framework (Rubin
1974). Here, you can imagine that each individual has a set of potential
outcomes under each possible exposure value. For example, if there are
two levels of exposure (exposed: 1 and unexposed: 0), we could have the
potential outcome under exposure (\(Y(1)\)) and the potential outcome
under no exposure (\(Y(0)\)) and look at the difference between these,
\(Y(1) - Y(0)\) to understand the impact on the exposure value on the
outcome, \(Y\). Of course, at any moment in time, only one of these
potential outcomes is observable, the potential outcome corresponding to
the exposure the individual actually experienced. Under certain
assumptions, we can borrow information from individuals who have
received different exposures to compare the average difference between
their observed outcomes. First, we assume that the causal question you
think you are answering is \emph{consistent} with the one you are
actually asking via your analysis. We likewise assume that the exposure
is well (and singly) defined. That is, there is only one definition of
``exposure'' and it is equally defined for all individuals under study
(the assumption is that there are \emph{not multiple versions of
exposure}). We also make the assumption that one individual's exposure
does not impact the outcome of any other individual (this is often
referred to as an assumption of \emph{no interference}). These first
three assumptions are referred as the
\emph{stable-unit-treatment-value-assumption} or SUTVA (Imbens and Rubin
2015). We assume that everyone has some chance of having each level of
the exposure (this assumption is often called \emph{positivity}). And
finally, we assume that the potential outcomes are independent of the
exposure value the individual happened to experience given the
covariate(s) \emph{that are adjusted for} in our modeling process (this
assumption is often referred to as \emph{exchangeability}) (Hernán
2012). We \emph{do} assume of course that the exposure itself may cause
the outcome, but we assume that the \emph{assignment} to a specific
exposure value for a given individual is independent of their outcome.
The easiest way to think about this is the best case scenario for
estimating causal effects where the exposure is \emph{randomly assigned}
to each individual, ensuring that this assumption is true without the
need to adjust for any other factors. In non-randomized settings, we
likely need to adjust for other factors to satisfy this independence.
The problem is identifying which factors are required, as adjusting for
all observed factors may not be appropriate (and some may even give you
the wrong effect). The purpose of this paper is to focus on the observed
covariates, \(Z\). Given you have three variables, an exposure, \(X\),
an outcome, \(Y\), and some measured factor, \(Z\), how do you decide
whether you should estimate the average treatment effect adjusting for
\(Z\)?

\hypertarget{causal-diagrams-primer}{%
\subsection{Causal diagrams primer}\label{causal-diagrams-primer}}

\begin{figure}

{\centering \includegraphics{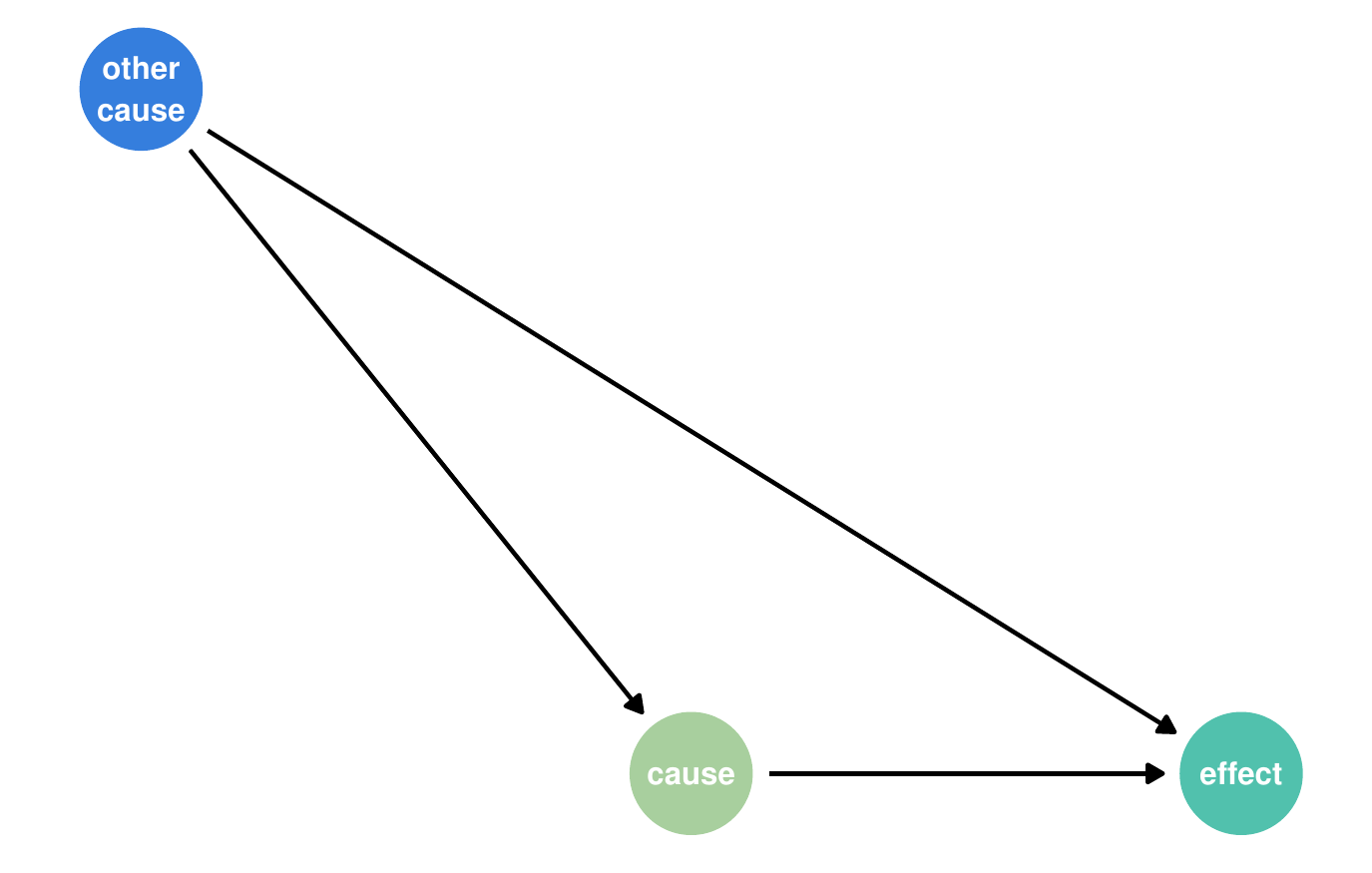}

}

\caption{\label{fig-dag}Example DAG. Here, there are three nodes
representing three factors: \texttt{cause}, \texttt{other\ cause}, and
\texttt{effect}. The arrows demonstrate the causal relationships between
these factors such that \texttt{cause} causes \texttt{effect} and
\texttt{other\ cause} causes both \texttt{cause} and \texttt{effect}.}

\end{figure}

Directed acyclic graphs (DAGs) are a mechanism used to communicate
causal relationships between factors. Factors are represented as
\emph{nodes} on the graphs, connected by directed \emph{edges} (arrows).
The edges point from causes to effects. The term \emph{acyclic} refers
to the fact that these graphs cannot have cycles. This is intuitive when
thinking about causes and effects, as a cycle would not be possible
without breaking the space-time continuum. DAGs are often used to
communicate proposed causal relationships between a set of factors. For
example, Figure~\ref{fig-dag} displays a DAG that suggests that the
\texttt{cause} causes \texttt{effect} and \texttt{other\ cause} causes
both \texttt{cause} and \texttt{effect}.

\hypertarget{causal-quartet}{%
\subsubsection{Causal quartet}\label{causal-quartet}}

\begin{figure}

{\centering \includegraphics{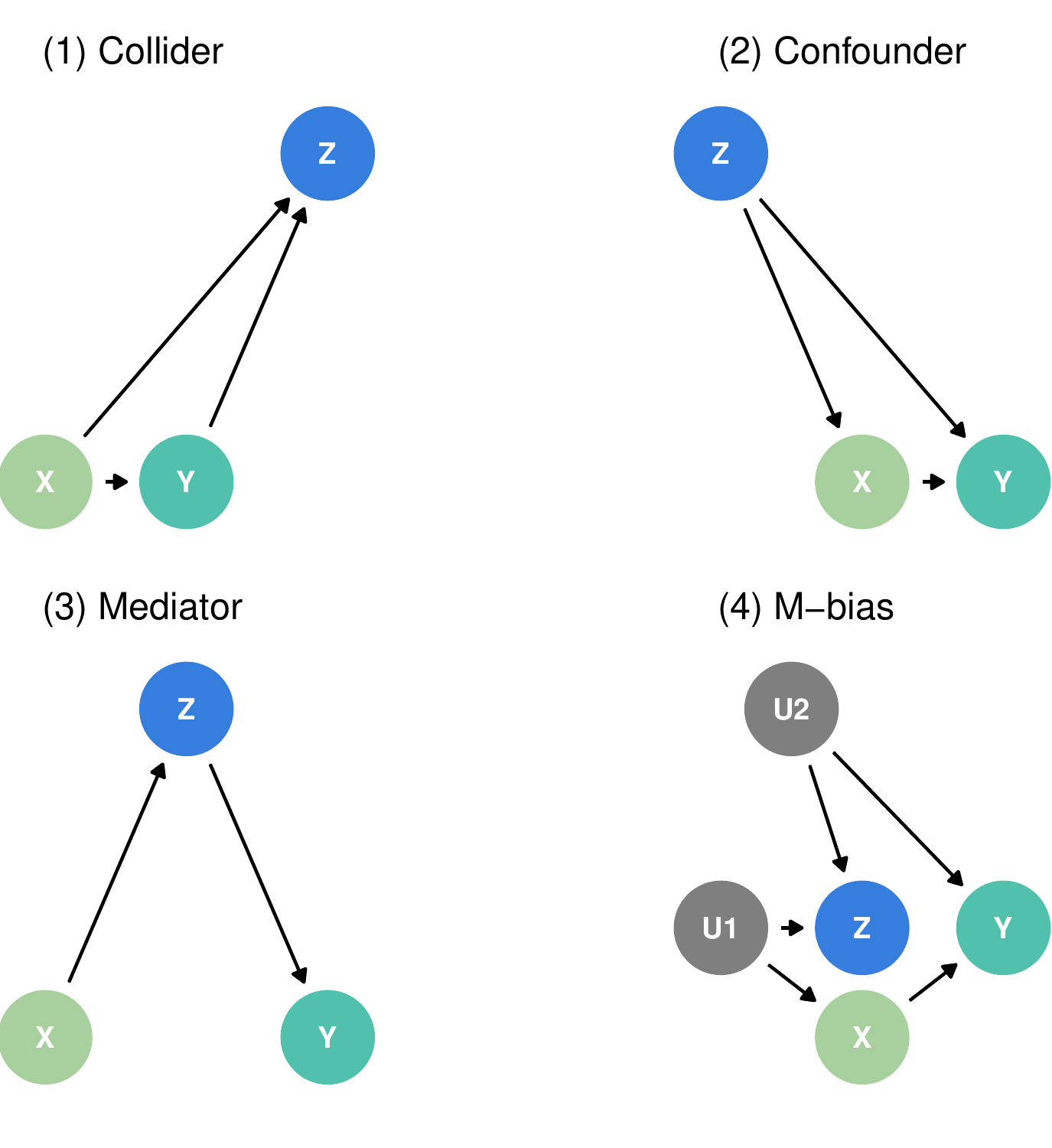}

}

\caption{\label{fig-1}Directed Acyclic Graphs describing the four data
generating mechanisms: (1) Collider (2) Confounder (3) Mediator (4)
M-Bias.}

\end{figure}

We propose the following four data generation mechanisms, summarized by
the equations below, as well as the DAGs displayed in
Figure~\ref{fig-1}. Here, \(X\) is presumed to be some continuous
exposure of interest, \(Y\) a continuous outcome, and \(Z\) a known,
measured factor. The M-Bias equation includes two additional, unmeasured
factors, \(U_1\) and \(U_2\).

\hypertarget{tbl-data-gen}{}
\begin{longtable}[]{@{}
  >{\raggedright\arraybackslash}p{(\columnwidth - 4\tabcolsep) * \real{0.1900}}
  >{\raggedright\arraybackslash}p{(\columnwidth - 4\tabcolsep) * \real{0.3900}}
  >{\raggedright\arraybackslash}p{(\columnwidth - 4\tabcolsep) * \real{0.4200}}@{}}
\caption{\label{tbl-data-gen}Causal terminology along with the data
generating mechanism for each of the four data sets included in the
causal quartet.}\tabularnewline
\toprule()
\begin{minipage}[b]{\linewidth}\raggedright
\hypertarget{technical-term}{%
\subsubsection{Technical term}\label{technical-term}}
\end{minipage} & \begin{minipage}[b]{\linewidth}\raggedright
\hypertarget{explanation}{%
\subsubsection{Explanation}\label{explanation}}
\end{minipage} & \begin{minipage}[b]{\linewidth}\raggedright
\hypertarget{data-generating-mechanism}{%
\subsubsection{Data generating
mechanism}\label{data-generating-mechanism}}
\end{minipage} \\
\midrule()
\endfirsthead
\toprule()
\begin{minipage}[b]{\linewidth}\raggedright
\hypertarget{technical-term}{%
\subsubsection{Technical term}\label{technical-term}}
\end{minipage} & \begin{minipage}[b]{\linewidth}\raggedright
\hypertarget{explanation}{%
\subsubsection{Explanation}\label{explanation}}
\end{minipage} & \begin{minipage}[b]{\linewidth}\raggedright
\hypertarget{data-generating-mechanism}{%
\subsubsection{Data generating
mechanism}\label{data-generating-mechanism}}
\end{minipage} \\
\midrule()
\endhead
(1) Collider & The exposure, \(X\), causes a factor, \(Z\), \emph{and}
the outcome, \(Y\), causes a factor, \(Z\). Adjusting for \(Z\) when
estimating the effect of \(X\) on \(Y\) would yield a biased result. &
\(X \sim N(0, 1)\)

\(Y = X + \varepsilon_y, \textrm{ }\varepsilon_y\sim N(0, 1)\)

\(Z = 0.45X + 0.77 Y + \varepsilon_z, \textrm{ }\varepsilon_z \sim N(0,1)\) \\
(2) Confounder & A factor, \(Z\), causes both the exposure, \(X\), and
the outcome, \(Y\). Failing to adjust for \(Z\) when estimating the
effect of \(X\) on \(Y\) would yield a biased result. &
\(Z \sim N(0, 1)\)

\(X = Z + \varepsilon_x,\textrm{ }\varepsilon_x\sim N(0, 1)\)

\(Y = 0.5X + Z + \varepsilon_y, \textrm{ }\varepsilon_y\sim N(0, 1)\) \\
(3) Mediator & An exposure, \(X\), causes a factor, \(Z\), which causes
the outcome, \(Y\). Adjusting for \(Z\) when estimating the effect of
\(X\) on \(Y\) would yield the \emph{direct effect,} not adjusting for
\(Z\) would yield the \emph{total effect} of \(X\) on \(Y\). The
\emph{direct effect} represents the relationship between \(X\) and \(Y\)
independent of any mediator, while the \emph{total effect} includes both
the direct effect and any indirect effects mediated by the potential
mediator. & \(X \sim N(0, 1)\)
\(Z = X + \varepsilon_z, \textrm{ }\varepsilon_z\sim N(0, 1)\)
\(Y = Z + \varepsilon_y, \textrm{ }\varepsilon_y\sim N(0, 1)\) \\
(4) M-Bias & There are two additional factors, \(U_1\) and \(U_2\). Both
cause \(Z\), \(U_1\) causes the exposure, \(X\), and \(U_2\) causes the
outcome, \(Y\). Adjusting for \(Z\) when estimating the effect of \(X\)
on \(Y\) will yield a biased result. & \(U_1 \sim N(0, 1)\)

\(U_2 \sim N(0, 1)\)

\(Z = 8 U_1 + U_2 + \varepsilon_z, \textrm{ }\varepsilon_z\sim N(0, 1)\)

\(X = U_1 + \varepsilon_x, \textrm{ }\varepsilon_x\sim N(0, 1)\)

\(Y = X + U_2 + \varepsilon_y, \textrm{ }\varepsilon_y\sim N(0, 1)\) \\
\bottomrule()
\end{longtable}

In each of these scenarios, a linear model fit to estimate the
relationship between \(X\) and \(Y\) with no further adjustment will
result in an expected \(\hat\beta\) coefficient of 1. Or, equivalently,
the expected estimated average treatment effect (\(\hat{\textrm{ATE}}\))
without adjusting for \(Z\) is 1. The correlation between \(X\) and the
additional known factor \(Z\) is also 0.70.

\begin{figure}

{\centering \includegraphics{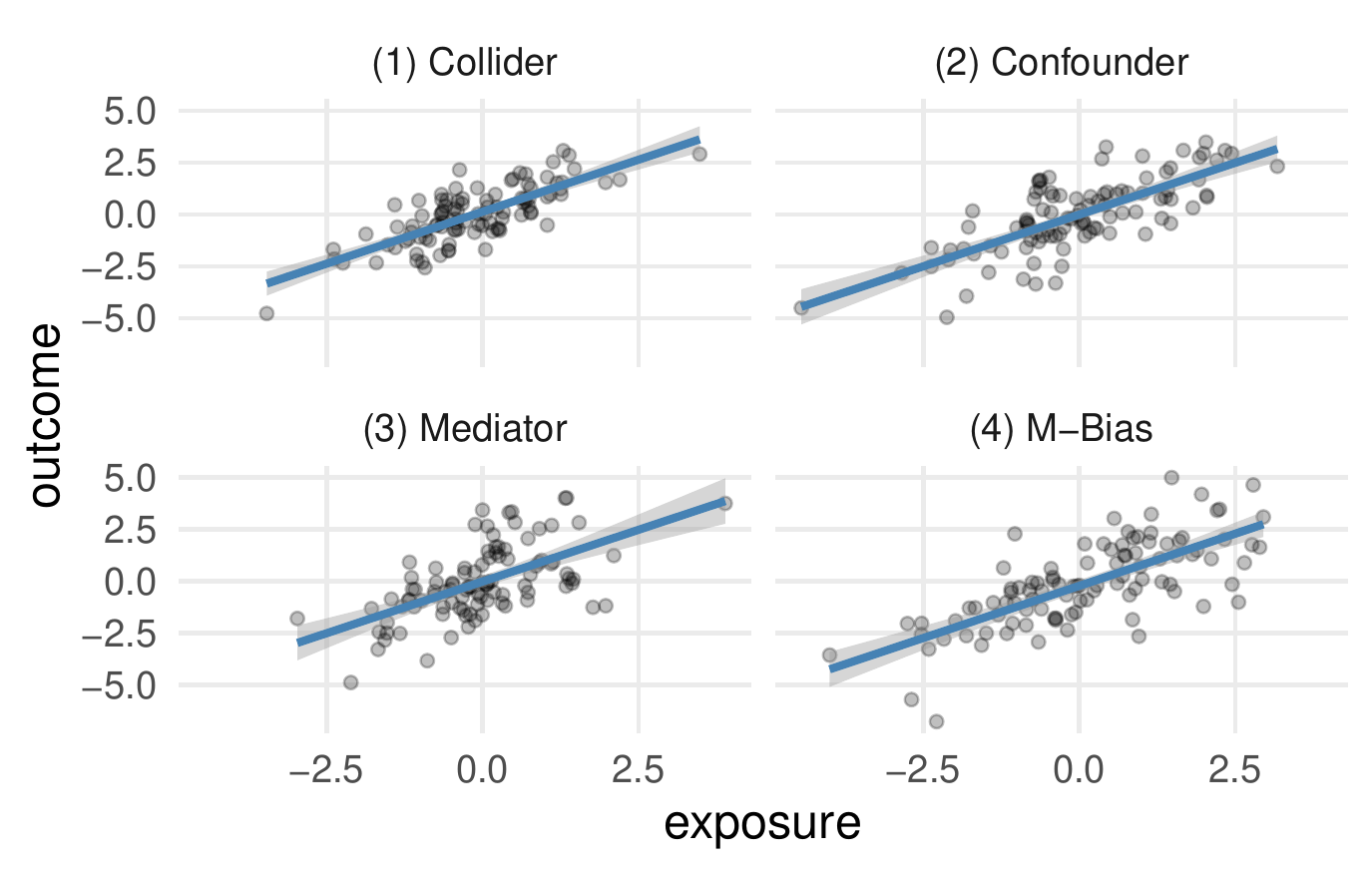}

}

\caption{\label{fig-2}100 points generated using the data generating
mechanisms specified (1) Collider (2) Confounder (3) Mediator (4)
M-Bias. The blue line displays a linear regression fit estimating the
relationship between X and Y; in each case, the slope is 1.}

\end{figure}

We have simulated 100 data points from each of the four mechanisms; we
display each in Figure~\ref{fig-2}. This set of figures demonstrates
that despite the very different data-generating mechanisms, there is no
clear way to determine the ``appropriate'' way to model the effect of
the exposure \(X\) and the outcome \(Y\) without additional information.
For example, the unadjusted models are displayed in Figure~\ref{fig-2},
showing a relationship between \(X\) and \(Y\) of 1. The unadjusted
models are the correct causal model for data-generating mechanisms (1)
and (4); however, it overstates the effect of \(X\) for data-generating
mechanism (2) and describes the total effect of \(X\) on \(Y\) for
data-generating mechanism (3), but not the direct effect
(Table~\ref{tbl-2}). Even examining the correlation between \(X\) and
the known factor \(Z\) does not help us determine whether adjusting for
\(Z\) is appropriate, as it is 0.7 in all cases (Table~\ref{tbl-3}). It
is commonly suggested when attempting to estimate causal effects using
observational data that the design step (i.e., selecting which variables
to adjust for) should be separate from the analysis step (i.e., fitting
an outcome model) (Rubin 2008). Even following the advice to choose
which variables to adjust for without examining any outcome data can
result in adjusting for factors that would lead to spurious estimates of
the causal effect, as seen here where the correlation between \(X\) and
\(Z\) is the same in every data set even though adjusting for \(Z\) is
sometimes not correct. Additionally, while it is not recommended to
choose which factors to adjust using the outcome variable (as this can
lead to increased Type 1 error and breaks the philosophical emulation of
a randomized trial), if we examine the correlation between \(Z\) and
\(Y\), we find that it is positive in all four cases as well.
Specifically, looking at the collider and confounder examples, the
correlation between \(Z\) and \(Y\) is approximately the same (0.8) in
the example data sets, and yet the confounder ought to be adjusted for
and the collider not.

Each of the four data sets described above are available for use in the
\texttt{quartets} R package (D'Agostino McGowan 2023). When using these
data sets in the classroom, potential real-world examples could be
assigned to the variables in line with the domain expertise of the
students. For example, if the course were taught to medical
professionals, the exposure, \(X\), could be sodium intake, the outcome,
\(Y\), systolic blood pressure, and a collider, \(Z\), urinary protein
excretion (Luque-Fernandez et al. 2019). See the \texttt{quartets}
package vignette titled \emph{``A collider example in a medical
context''} for an example of a lesson plan using this framework
(D'Agostino McGowan 2023).

\hypertarget{tbl-2}{}
\begin{longtable}[]{@{}
  >{\raggedright\arraybackslash}p{(\columnwidth - 4\tabcolsep) * \real{0.3457}}
  >{\raggedright\arraybackslash}p{(\columnwidth - 4\tabcolsep) * \real{0.3457}}
  >{\raggedright\arraybackslash}p{(\columnwidth - 4\tabcolsep) * \real{0.2963}}@{}}
\caption{\label{tbl-2}Correct causal models and causal effects for each
data-generating mechanism. The notation \(X ; Z\) implies that we should
adjust for \(Z\) when estimating the causal effect. In other words, for
the confounder data generating mechanism and direct effect mediator
model, the potential outcomes are independent of exposure given the
observed factor \(Z\).}\tabularnewline
\toprule()
\begin{minipage}[b]{\linewidth}\raggedright
Data generating mechanism
\end{minipage} & \begin{minipage}[b]{\linewidth}\raggedright
Correct causal model
\end{minipage} & \begin{minipage}[b]{\linewidth}\raggedright
Correct causal effect
\end{minipage} \\
\midrule()
\endfirsthead
\toprule()
\begin{minipage}[b]{\linewidth}\raggedright
Data generating mechanism
\end{minipage} & \begin{minipage}[b]{\linewidth}\raggedright
Correct causal model
\end{minipage} & \begin{minipage}[b]{\linewidth}\raggedright
Correct causal effect
\end{minipage} \\
\midrule()
\endhead
(1) Collider & Y \textasciitilde{} X & 1 \\
(2) Confounder & Y \textasciitilde{} X ; Z & 0.5 \\
(3) Mediator & Direct effect: Y \textasciitilde{} X ; Z

Total Effect: Y \textasciitilde{} X & Direct effect: 0

Total effect: 1 \\
(4) M-Bias & Y \textasciitilde{} X & 1 \\
\bottomrule()
\end{longtable}

\newpage

\hypertarget{tbl-3}{}
\begin{table}
\caption{\label{tbl-3}Estimated average treatment effects under each data generating mechanism
with and without adjustment for Z as well as the correlation between X
and Z. }\tabularnewline

\centering
\begin{tabular}{lrrr}
\toprule
Data generating mechanism & \makecell[c]{$\hat{\textrm{ATE}}$\\not adjusting for Z} & \makecell[c]{$\hat{\textrm{ATE}}$\\adjusting for Z} & \makecell[c]{Correlation of\\X and Z}\\
\midrule
(1) Collider & 1 & 0.55 & 0.7\\
(2) Confounder & 1 & 0.50 & 0.7\\
(3) Mediator & 1 & 0.00 & 0.7\\
(4) M-Bias & 1 & 0.88 & 0.7\\
\bottomrule
\end{tabular}
\end{table}

\hypertarget{the-solution}{%
\subsubsection{The solution}\label{the-solution}}

Here we have demonstrated that when presented with an exposure, outcome,
and some measured factors, statistics alone, whether summary statistics
or data visualizations, are insufficient to determine the appropriate
causal estimate. Analysts need additional information about the data
generating mechanism to draw the correct conclusions. While knowledge of
the data generating process is necessary to estimate the correct causal
effect in each of the cases presented, an analyst can take steps to make
mistakes such as those shown here less likely. The first is discussing
understood mechanisms with content matter experts before estimating
causal effects. Drawing the proposed relationships via causal diagrams
such as the directed acyclic graphs shown in Figure~\ref{fig-1} before
calculating any statistical quantities can help the analyst ensure they
are only adjusting for factors that meet the ``backdoor criterion,''
that is, adjusting for only factors that close all backdoor paths
between the exposure and outcome of interest (Pearl 2000).

\begin{figure}

\begin{minipage}[t]{0.50\linewidth}

{\centering 

\raisebox{-\height}{

\includegraphics{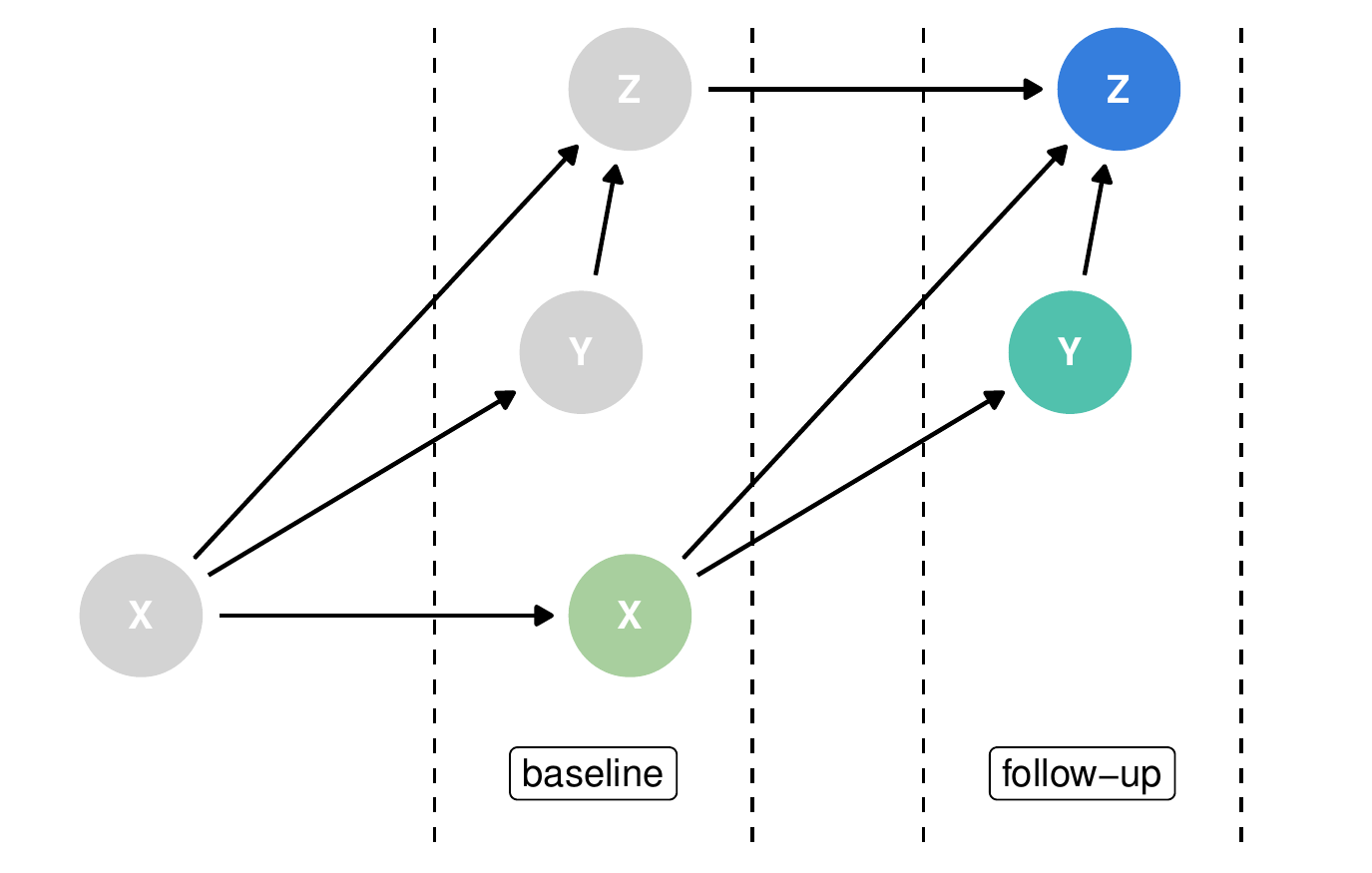}

}

}

\subcaption{\label{fig-3-1}Adjusting for \(Z\) as shown here would
induce collider bias.}
\end{minipage}%
\begin{minipage}[t]{0.50\linewidth}

{\centering 

\raisebox{-\height}{

\includegraphics{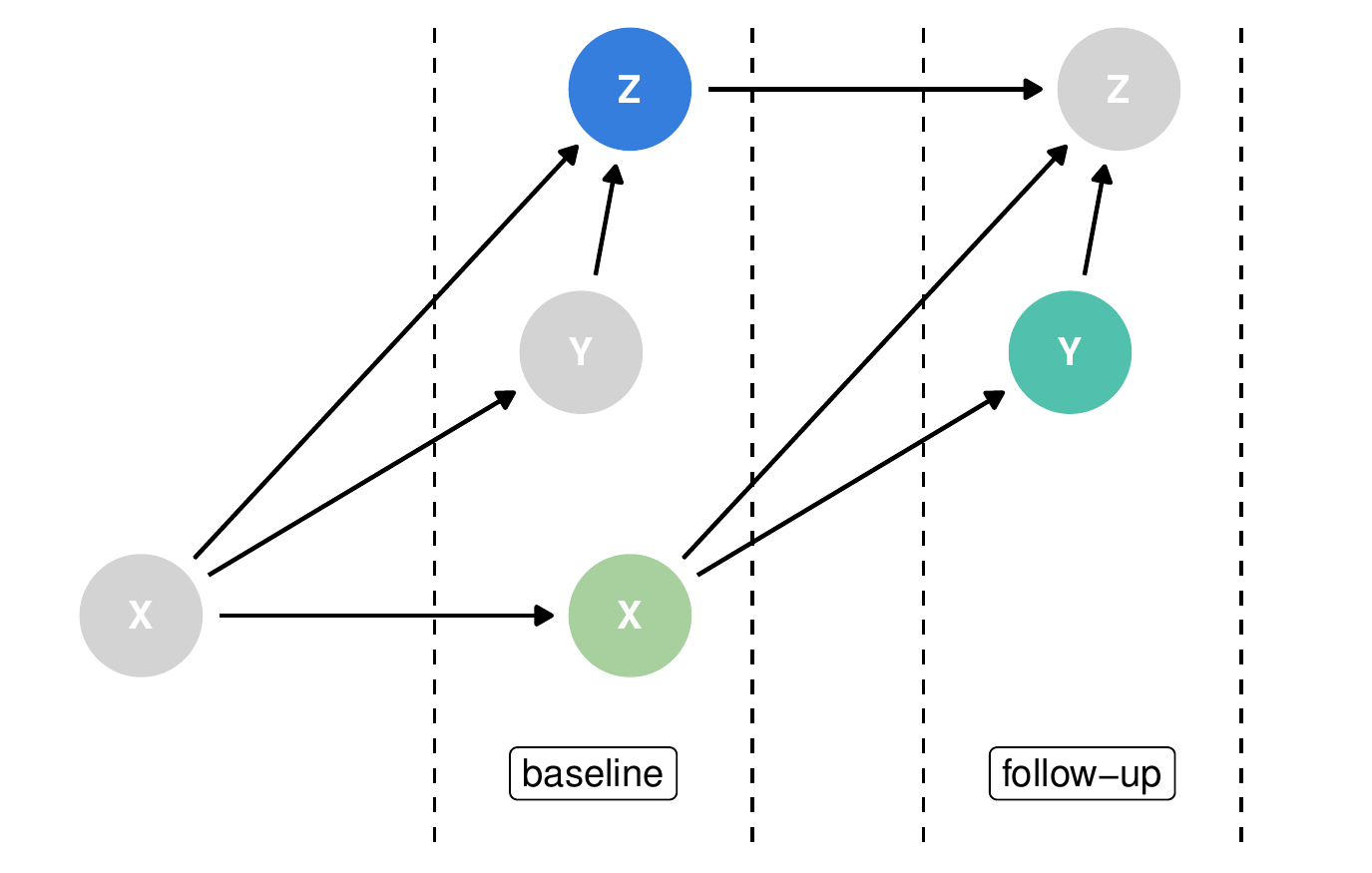}

}

}

\subcaption{\label{fig-3-2}Adjusting for this pre-exposure \(Z\) as
shown here would \textbf{not} induce collider bias.}
\end{minipage}%

\caption{\label{fig-3}Time-ordered collider DAG (with time increasing
from left to right) where each factor is measured twice. \(X\) is the
exposure, \(Y\) is the outcome, and \(Z\) is the measured factor. The
highlighted \(Z\) node indicates which time point is being adjusted for
when estimating the average treatment effect of the highlighted \(X\) on
the highlighted \(Y\)}

\end{figure}

Absent subject matter expertise, the analyst can at least consider the
time ordering of the available factors. Fundamental principles of causal
inference dictate that the exposure of interest must precede the outcome
of interest to establish a causal relationship plausibly. In addition,
to account for potential confounding, any covariates adjusted for in the
analysis must precede the exposure in time. Including this additional
timing information would omit the potential for two of the three
misspecified models above (Table~\ref{tbl-data-gen}, the ``collider''
and the ``mediator'') as the former would demonstrate that the factor
\(Z\) falls after both the exposure and outcome and the latter would
show that the factor \(Z\) falls between the exposure and the outcome in
time. For example, if we drew the second panel of Figure~\ref{fig-1}
(the Collider) as a time ordered DAG, we would see something like
Figure~\ref{fig-3}. If we carefully adjust only for factors that are
measured pre-exposure, we would not induce the bias we see in
Table~\ref{tbl-3} (Figure~\ref{fig-3-2}). The \emph{causal quartet} data
sets are accompanied by a set of four data sets with time-varying
measures for each of the factors, \(X\), \(Y\), and \(Z\), generated
under the same data generating mechanisms. Here, as long as a
pre-exposure measure of \(Z\) is adjusted for, the correct causal effect
is estimated in all scenarios except M-Bias (Table~\ref{tbl-4}). These
data sets also serve the useful pedagogical purpose that since time is
included in the information provided there are particular DAGs that are
always incorrect. For example, an arrow between \(Z\) at follow-up and
the \(X\) at baseline is impossible, since, as far as the authors are
aware, time travel is not possible. That is, factors in the future
cannot cause effects in the past.

\hypertarget{tbl-4}{}
\begin{table}
\caption{\label{tbl-4}Coefficients for the exposure under each data generating mechanism
depending on the model fit as well as the correlation between X and Z. }\tabularnewline

\centering
\begin{tabular}{lrrr}
\toprule
Data generating mechanism & \makecell[c]{ATE\\not adjusting for\\pre-exposure Z} & \makecell[c]{ATE\\adjusting for\\pre-exposure Z} & Correct causal effect\\
\midrule
(1) Collider & 1 & 1.00 & 1.0\\
(2) Confounder & 1 & 0.50 & 0.5\\
(3) Mediator & 1 & 1.00 & 1.0\\
(4) M-Bias & 1 & 0.88 & 1.0\\
\bottomrule
\end{tabular}
\end{table}

Adjusting for only pre-exposure factors is widely recommended. The only
exception is when a known confounder is only measured after the exposure
in a particular data analysis, in which case some experts recommend
adjusting for it. Still, even then, caution is advised (Groenwold,
Palmer, and Tilling 2021). Many causal inference methodologists would
recommend conditioning on \emph{all} measured pre-exposure factors
(Rosenbaum 2002; Rubin 2009, 2008; Rubin and Thomas 1996). Including
timing information alone (and thus adjusting for all pre-exposure
factors) does not preclude one from mistakenly fitting the adjusted
model under the fourth data generating mechanism (M-bias), as \(Z\) can
fall temporally before \(X\) and \(Y\) and still induce bias. It has
been argued, however, that this strict M-bias (e.g., as in
Table~\ref{tbl-data-gen} where \(U_1\) and \(U_2\) have no relationship
with each other and \(Z\) has no relationship with \(X\) or \(Y\) other
than via \(U_1\) and \(U_2\)) is very rare in most practical settings
(Liu et al. 2012; Rubin 2009; Gelman 2011). Indeed, even theoretical
results have demonstrated that bias induced by this data generating
mechanism is sensitive to any deviations from this form (Ding and
Miratrix 2015).

\hypertarget{discussion}{%
\subsection{Discussion}\label{discussion}}

In the spirit of Anscombe's Quartet, small data sets created to
demonstrate a key concept akin to those we introduce here have been used
for a wide variety of data analytic problems. Recent examples include an
extension of the original idea proposed by Anscombe called the
``Datasaurus Dozen'' (Matejka and Fitzmaurice 2017), an exploration of
varying interaction effects (Rohrer and Arslan 2021), a quartet of model
types fit to the same data that yield the same performance metrics but
fit very different underlying mechanisms (Biecek, Baniecki, and
Krzyznski 2023), and a set of conceptual causal quartets that highlight
the impact of treatment heterogeneity on the average treatment effect
(Gelman, Hullman, and Kennedy 2023). While similar in name, the
conceptual causal quartets are different from what we present here as
they provide excellent insight into how variation in a treatment effect
/ treatment heterogeneity can impact an average treatment effect (by
plotting the latent true causal effect). We believe both sets provide
important and complementary understanding for data analysis
practitioners.

We have presented four example data sets demonstrating the importance of
understanding the data-generating mechanism when attempting to answer
causal questions. These data indicate that more than statistical
summaries and visualizations are needed to provide insight into the
underlying relationship between the variables. An understanding or
assumption of the data-generating mechanism is required to capture
causal relationships correctly. These examples underscore the
limitations of relying solely on statistical tools in data analyses and
highlight the crucial role of domain-specific knowledge. Moreover, they
emphasize the importance of considering the timing of factors when
deciding what to adjust for.

\hypertarget{references}{%
\subsection{References}\label{references}}

\hypertarget{refs}{}
\begin{CSLReferences}{1}{0}
\leavevmode\vadjust pre{\hypertarget{ref-anscombe1973graphs}{}}%
Anscombe, Francis J. 1973. {``Graphs in Statistical Analysis.''}
\emph{The American Statistician} 27 (1): 17--21.

\leavevmode\vadjust pre{\hypertarget{ref-biecek2023performance}{}}%
Biecek, Przemyslaw, Hubert Baniecki, and Mateusz Krzyznski. 2023.
{``Performance Is Not Enough: A Story of the Rashomon's Quartet.''}
\emph{arXiv Preprint arXiv:2302.13356}.

\leavevmode\vadjust pre{\hypertarget{ref-quartet}{}}%
D'Agostino McGowan, Lucy. 2023. \emph{Quartets: Datasets to Help Teach
Statistics}.

\leavevmode\vadjust pre{\hypertarget{ref-ding2015adjust}{}}%
Ding, Peng, and Luke W Miratrix. 2015. {``To Adjust or Not to Adjust?
Sensitivity Analysis of m-Bias and Butterfly-Bias.''} \emph{Journal of
Causal Inference} 3 (1): 41--57.

\leavevmode\vadjust pre{\hypertarget{ref-gelman2011causality}{}}%
Gelman, Andrew. 2011. {``Causality and Statistical Learning.''}
University of Chicago Press Chicago, IL.

\leavevmode\vadjust pre{\hypertarget{ref-gelman2023causal}{}}%
Gelman, Andrew, Jessica Hullman, and Lauren Kennedy. 2023. {``Causal
Quartets: Different Ways to Attain the Same Average Treatment Effect.''}
\emph{arXiv Preprint arXiv:2302.12878}.

\leavevmode\vadjust pre{\hypertarget{ref-groenwold2021adjust}{}}%
Groenwold, Rolf HH, Tom M Palmer, and Kate Tilling. 2021. {``To Adjust
or Not to Adjust? When a {`Confounder'} Is Only Measured After
Exposure.''} \emph{Epidemiology (Cambridge, Mass.)} 32 (2): 194.

\leavevmode\vadjust pre{\hypertarget{ref-hernan2012beyond}{}}%
Hernán, Miguel A. 2012. {``Beyond Exchangeability: The Other Conditions
for Causal Inference in Medical Research.''} \emph{Statistical Methods
in Medical Research}. Sage Publications Sage UK: London, England.

\leavevmode\vadjust pre{\hypertarget{ref-imbens2015causal}{}}%
Imbens, Guido W, and Donald B Rubin. 2015. \emph{Causal Inference in
Statistics, Social, and Biomedical Sciences}. Cambridge University
Press.

\leavevmode\vadjust pre{\hypertarget{ref-liu2012implications}{}}%
Liu, Wei, M Alan Brookhart, Sebastian Schneeweiss, Xiaojuan Mi, and Soko
Setoguchi. 2012. {``Implications of m Bias in Epidemiologic Studies: A
Simulation Study.''} \emph{American Journal of Epidemiology} 176 (10):
938--48.

\leavevmode\vadjust pre{\hypertarget{ref-luque2019educational}{}}%
Luque-Fernandez, Miguel Angel, Michael Schomaker, Daniel
Redondo-Sanchez, Maria Jose Sanchez Perez, Anand Vaidya, and Mireille E
Schnitzer. 2019. {``Educational Note: Paradoxical Collider Effect in the
Analysis of Non-Communicable Disease Epidemiological Data: A
Reproducible Illustration and Web Application.''} \emph{International
Journal of Epidemiology} 48 (2): 640--53.

\leavevmode\vadjust pre{\hypertarget{ref-matejka2017same}{}}%
Matejka, Justin, and George Fitzmaurice. 2017. {``Same Stats, Different
Graphs: Generating Datasets with Varied Appearance and Identical
Statistics Through Simulated Annealing.''} In \emph{Proceedings of the
2017 CHI Conference on Human Factors in Computing Systems}, 1290--94.

\leavevmode\vadjust pre{\hypertarget{ref-pearl2000causality}{}}%
Pearl, Judea. 2000. \emph{Causality: Models, Reasoning, and Inference}.
Cambridge University Press.

\leavevmode\vadjust pre{\hypertarget{ref-rohrer2021precise}{}}%
Rohrer, Julia M, and Ruben C Arslan. 2021. {``Precise Answers to Vague
Questions: Issues with Interactions.''} \emph{Advances in Methods and
Practices in Psychological Science} 4 (2): 25152459211007368.

\leavevmode\vadjust pre{\hypertarget{ref-rosenbaumconstructing}{}}%
Rosenbaum, PR. 2002. {``Constructing Matched Sets and Strata.
Observational Studies.''} New York, Springer-Verlag.

\leavevmode\vadjust pre{\hypertarget{ref-rubin1974estimating}{}}%
Rubin, Donald B. 1974. {``Estimating Causal Effects of Treatments in
Randomized and Nonrandomized Studies.''} \emph{Journal of Educational
Psychology} 66 (5): 688.

\leavevmode\vadjust pre{\hypertarget{ref-rubin2008objective}{}}%
---------. 2008. {``For Objective Causal Inference, Design Trumps
Analysis.''}

\leavevmode\vadjust pre{\hypertarget{ref-rubin2009should}{}}%
---------. 2009. {``Should Observational Studies Be Designed to Allow
Lack of Balance in Covariate Distributions Across Treatment Groups?''}
\emph{Statistics in Medicine} 28 (9): 1420--23.

\leavevmode\vadjust pre{\hypertarget{ref-rubin1996matching}{}}%
Rubin, Donald B, and Neal Thomas. 1996. {``Matching Using Estimated
Propensity Scores: Relating Theory to Practice.''} \emph{Biometrics},
249--64.

\end{CSLReferences}

\hypertarget{appendix}{%
\subsection{Appendix}\label{appendix}}

R code to generate the tables and figures:

\begin{Shaded}
\begin{Highlighting}[]
\FunctionTok{library}\NormalTok{(tidyverse)}
\CommentTok{\# install.packages("quartets")}
\FunctionTok{library}\NormalTok{(quartets)}

\DocumentationTok{\#\# Figure 2}

\FunctionTok{ggplot}\NormalTok{(causal\_quartet, }\FunctionTok{aes}\NormalTok{(}\AttributeTok{x =}\NormalTok{ exposure, }\AttributeTok{y =}\NormalTok{ outcome)) }\SpecialCharTok{+}
  \FunctionTok{geom\_point}\NormalTok{(}\AttributeTok{alpha =} \FloatTok{0.25}\NormalTok{) }\SpecialCharTok{+} 
  \FunctionTok{geom\_smooth}\NormalTok{(}
    \AttributeTok{method =} \StringTok{"lm"}\NormalTok{, }
    \AttributeTok{formula =} \StringTok{"y \textasciitilde{} x"}\NormalTok{, }
    \AttributeTok{linewidth =} \FloatTok{1.1}\NormalTok{, }
    \AttributeTok{color =} \StringTok{"steelblue"}
\NormalTok{  ) }\SpecialCharTok{+}
  \FunctionTok{facet\_wrap}\NormalTok{(}\SpecialCharTok{\textasciitilde{}}\NormalTok{dataset)}
\DocumentationTok{\#\# Table 3}

\NormalTok{tbl\_3 }\OtherTok{\textless{}{-}}\NormalTok{ causal\_quartet }\SpecialCharTok{|\textgreater{}}
  \FunctionTok{nest\_by}\NormalTok{(dataset) }\SpecialCharTok{|\textgreater{}}
  \FunctionTok{mutate}\NormalTok{(}\AttributeTok{ate\_x =} \FunctionTok{coef}\NormalTok{(}\FunctionTok{lm}\NormalTok{(outcome }\SpecialCharTok{\textasciitilde{}}\NormalTok{ exposure, }\AttributeTok{data =}\NormalTok{ data))[}\DecValTok{2}\NormalTok{],}
         \AttributeTok{ate\_xz =} \FunctionTok{coef}\NormalTok{(}\FunctionTok{lm}\NormalTok{(outcome }\SpecialCharTok{\textasciitilde{}}\NormalTok{ exposure }\SpecialCharTok{+}\NormalTok{ covariate, }\AttributeTok{data =}\NormalTok{ data))[}\DecValTok{2}\NormalTok{],}
         \AttributeTok{cor =} \FunctionTok{cor}\NormalTok{(data}\SpecialCharTok{$}\NormalTok{exposure, data}\SpecialCharTok{$}\NormalTok{covariate)) }\SpecialCharTok{|\textgreater{}}
  \FunctionTok{select}\NormalTok{(}\SpecialCharTok{{-}}\NormalTok{data, dataset)}
\DocumentationTok{\#\# Table 4}

\NormalTok{tbl\_4 }\OtherTok{\textless{}{-}}\NormalTok{ causal\_quartet\_time }\SpecialCharTok{|\textgreater{}}
  \FunctionTok{nest\_by}\NormalTok{(dataset) }\SpecialCharTok{|\textgreater{}}
  \FunctionTok{mutate}\NormalTok{(}\AttributeTok{ate\_x =} 
           \FunctionTok{coef}\NormalTok{(}
               \FunctionTok{lm}\NormalTok{(outcome\_followup }\SpecialCharTok{\textasciitilde{}}\NormalTok{ exposure\_baseline, }\AttributeTok{data =}\NormalTok{ data)}
\NormalTok{             )[}\DecValTok{2}\NormalTok{],}
         \AttributeTok{ate\_xz =} 
             \FunctionTok{coef}\NormalTok{(}
               \FunctionTok{lm}\NormalTok{(outcome\_followup }\SpecialCharTok{\textasciitilde{}}\NormalTok{ exposure\_baseline }\SpecialCharTok{+}\NormalTok{ covariate\_baseline, }
                  \AttributeTok{data =}\NormalTok{ data)}
\NormalTok{             )[}\DecValTok{2}\NormalTok{]) }\SpecialCharTok{|\textgreater{}}
  \FunctionTok{bind\_cols}\NormalTok{(}\FunctionTok{tibble}\NormalTok{(}\AttributeTok{truth =} \FunctionTok{c}\NormalTok{(}\DecValTok{1}\NormalTok{, }\FloatTok{0.5}\NormalTok{, }\DecValTok{1}\NormalTok{, }\DecValTok{1}\NormalTok{))) }\SpecialCharTok{|\textgreater{}}
  \FunctionTok{select}\NormalTok{(}\SpecialCharTok{{-}}\NormalTok{data, dataset)}
\end{Highlighting}
\end{Shaded}

\end{document}